# Causal Machine Learning Analysis of Empirical Relative Biological Effectiveness (RBE) for Mandible Osteoradionecrosis in Head and Neck Cancer Radiotherapy

**Authors:** Jingyuan Chen, PhD[1], Zhong Liu, PhD[2], Yunze Yang, PhD[3], Olivia M. Muller, MD[4], Zhengliang Liu, MS[5], Tianming Liu, PhD[5], Lei Zeng, MD[1,6], Robert L, Foote, MD[7], Daniel J, Ma, MD[7], Samir H, Patel, MD[1], Wei Liu, PhD[1]

[1]Department of Radiation Oncology, Mayo Clinic, Phoenix, AZ 85054, USA

[2]Institute of Western China Economic Research, Southwestern University of Finance and Economics, Chengdu, Sichuan 611130, China

[3]Department of Radiation Oncology, the University of Miami, FL 33136, USA

[4]Department of Dental Specialties, Mayo Clinic Rochester, Rochester, MN, 55905, USA

[5]Department of Computer Science, University of Georgia, Athens, GA, 30602, USA

[6]Department of Oncology, The Second Affiliated Hospital, Jiangxi Medical College, Nanchang University, Nanchang, 330006, China

[7]Department of Radiation Oncology, Mayo Clinic, Rochester, MN, 55905, USA

Corresponding author: Wei Liu, PhD, Professor of Radiation Oncology, Department of Radiation Oncology, Mayo Clinic Arizona; e-mail: Liu.Wei@mayo.edu.

**Conflicts of Interest Disclosure Statement**

No

**Funding Statement**

This research was supported by NIH/BIBIB R01EB293388, by NIH/NCI R01CA280134, by the Eric & Wendy Schmidt Fund for AI Research & Innovation, and by the Kemper Marley Foundation.

**Ethical Approval**

This study was approved by Mayo Clinic Arizona institutional review board (IRB#: 24-011106).

**Data Availability Statement**

The data analyzed during the current study are not publicly available due to patient privacy concerns and institutional policies regarding protected health information (PHI). However, de-identified data that support the findings of this study are available from the corresponding author upon reasonable request and with appropriate institutional review board (IRB) approval.

**Acknowledgments**

This research was supported by NIH/BIBIB R01EB293388, by NIH/NCI R01CA280134, by the Eric & Wendy Schmidt Fund for AI Research & Innovation, and by the Kemper Marley Foundation.

# Abstract

**Background and Aims:**

Osteoradionecrosis (ORN) of the mandible is one of the most severe adverse events (AEs) for head and neck (H&N) cancer radiotherapy. Previous retrospective investigations on real-world data relied heavily on conventional statistical models that primarily elucidate correlation rather than establishing causal relationships. Through the novel causal machine learning method, we aim to obtain empirical relative biological effectiveness (RBE) for mandible ORN in head and neck (H&N) cancer patients treated with pencil-beam-scanning proton therapy (PBSPT).

**Methods:**

1,266 H&N cancer patients were included: 335 patients treated by PBSPT and 931 patients treated by volumetric-modulated arc therapy (VMAT). We use 1:1 case-matching based on propensity scores to minimize the imbalance in clinical factors between patients treated with PBSPT and VMAT. The bias test of standardized mean differences (SMD) was applied on the case-matched patient cohorts. The causal machine learning method, causal forest (CF), was adopted to investigate the causal effects between dosimetric factors and the incidence of ORN. The dose volume constraints (DVCs) for VMAT and PBSPT were derived when the critical volumes of the derived DVCs lead to the largest average causal effect (ATE). RBE values were further empirically derived based on tolerance curves formed from the critical volumes of the derived DVCs. This was accomplished by comparing the equivalent constraint doses against the actual physical doses of PBSPT. To rigorously account for statistical variability in the RBE estimates, a bootstrap resampling method was applied to generate confidence intervals, thereby quantifying the uncertainty in the analysis.

**Results:**

335 VMAT patients were case-matched to 335 PBSPT patients; however, standardized mean bias analysis revealed persistent covariate imbalances within each group, indicating residual confounding influence. Using CF modeling, we identified DVCs of mandible for ORN and found that PBSPT had lower critical volumes than those of VMAT, leading to empirical RBE exceeding 1.1 in the moderate dose range (1.61 at 40 Gy[RBE=1.1], 1.30 at 50 Gy, and 1.13 at 60 Gy).

**Conclusion:**

This study presents a novel application of causal machine learning to evaluate mandible ORN in radiotherapy, identifying DVCs linked to the strongest causal effects and deriving empirical RBEs from equivalent constraint dose analysis based on the tolerance curve. The results indicate that proton RBE may significantly exceed 1.1 in the moderate dose range (40–60 Gy[RBE=1.1]), underscoring the importance of incorporating the variable RBE into PBSPT treatment planning to mitigate the risk of ORN.

# Introduction

Adverse events (AEs) commonly arise following head and neck (H&N) cancer radiotherapy due to the numerous adjacent organs-at-risk (OARs)(1-3). These toxicities increase the supportive interventions and markedly diminish the quality of life (QoL) of patients(4-8). Among the various treatment-related toxicities, mandible osteoradionecrosis (ORN) stands out as one of the most serious outcomes(9-12). Volumetric-modulated arc therapy (VMAT) and pencil-beam-scanning proton therapy (PBSPT) are two cutting-edge approaches in external radiotherapy(13-15). VMAT delivers highly targeted radiation distributions through one or several rotating arcs(16). In contrast, PBSPT represents the latest proton therapy(17-20). The key advantage of proton therapy lies in its characteristic finite energy deposition range (known as the Bragg Peak), after which radiation dose essentially stops(21,22). This unique property allows PBSPT to conform radiation more precisely to treatment targets while significantly reducing exposure to surrounding healthy organs and tissues(23,24).

However, the Bragg Peak not only brings dosimetric benefits, but also the challenge of relative biological effectiveness (RBE) in PBSPT(25,26). The sharp energy deposition ranges near the distal edge of the Bragg Peak elevate linear energy transfer (LET), making the biological effectiveness of PBSPT dependent not only on the physical dose but also on LET (27-30). A uniform RBE value of 1.1 is typically applied to account for the increased biological killing power of protons compared to photons in clinical practice(31,32). Nevertheless, experimental studies using cell lines suggest that RBE is not constant and can deviate from 1.1 under certain conditions(28,33). Notably, a higher RBE associated with brain necrosis was reported based on normal tissue complication probability (NTCP) modeling and tolerance dose comparison (34,35). Recently, Yang et al.(11) found an evaluated RBE between VMAT and PBSPT at moderate doses

(between 40 and 60 Gy[RBE=1.1]) in a retrospective study of ORN. These retrospective studies could provide patient outcome-based empirical RBEs, which are valuable for clinical applications.

However, almost all the patient outcome studies reported so far in radiotherapy have been predictive modeling, and therefore the conclusions drawn are correlational rather than causal. In real-world data, the bias introduced by patients' clinical factors are difficult to control by traditional statistical methods(36,37). Causal machine learning (Causal ML) methods offer a powerful framework to address this challenge(38,39). By controlling the bias from the unbalanced clinical factors, Causal ML enables the estimation of the causal effect of dosimetric factors, also known as the treatment effect(39-42).

In this study, we investigated RBE based on the causal effect between dosimetric factors and the incidence of mandible ORN, using patient outcome data of H&N cancer patients treated with PBSPT and VMAT at our institutions. We illustrate the inherent bias in our real-world data even after case matching, which may lead to biased results from the traditional statistical methods. We adopt the Causal Forest (CF) method, a nonparametric, forest-based approach designed to estimate the causal treatment effects of the dosimetric factors(42,43). Dose volume constraints (DVCs) for mandible ORN were obtained respectively for VMAT and PBSPT to achieve the largest causal effect. Empirical RBE values for PBSPT were estimated by determining the ratios between its equivalent constraint doses and corresponding physical doses of PBSPT. To the best of our knowledge, this work represents one of the first causality studies to analyze empirical RBEs.

## Methods and Materials

**Study Population**

This analysis included head and neck cancer patients who received definitive chemoradiotherapy at Mayo Clinic in Rochester and Arizona from April 2013 through August 2019, ensuring a minimum follow-up period exceeding 24 months. Eligible participants were those with pathologically confirmed malignancies who underwent curative-intent treatment using either volumetric modulated arc therapy (VMAT) or pencil beam scanning proton therapy (PBSPT). The cohort comprised 1,266 individuals without restrictions based on demographic factors including gender, age, ethnicity, or body weight. Treatment allocation between modalities (931 receiving VMAT versus 335 receiving PBSPT) was not based on predetermined clinical criteria.

Inclusion criteria specified: (1) fractional doses ranging from 1.2 Gy[RBE] to 2.2 Gy[RBE] per fraction; (2) minimum prescribed dose of 60 Gy[RBE] delivered to primary tumor volumes; and (3) for re-treatment cases, inclusion only when mandible exposure from re-treatment was minimal or when ORN developed prior to re-treatment.

Dosimetric notation followed standard conventions: PBSPT doses incorporated a relative biological effectiveness factor of 1.1, while VMAT doses reflected physical measurements (RBE=1.0). Clinical parameters collected from our institutional registry included age, gender, tumor stage, concurrent chemotherapy, hypertension and diabetes diagnoses, dental extraction history, smoking history, and current smoking status. Institutional review board authorization was obtained for this investigation (IRB: 24-011006).

**Treatment Plan and diagnosis of Osteoradionecrosis (ORN)**

Treatment planning for both VMAT and PBSPT utilized Eclipse™ software (Varian Medical Systems, Palo Alto, CA) on the simulation computed tomography of patients. Every plan was optimized to adhere to our institution's DVCs, whenever possible. More details about treatment plans are presented in the Supplementary materials.

Experienced clinicians diagnosed ORN through comprehensive assessments: direct clinical assessment revealing exposed bone, diagnostic imaging including panoramic radiography (Panorex), computed tomography (CT), magnetic resonance (MR), and positron emission tomography (PET), and/or tissue analysis from surgical procedures such as bone removal or jaw resection.

**Table 1** Characteristics for patient cohort from VMAT and PBSPT with and without ORN

|  | Total | Photon | | Proton | |
|---|---|---|---|---|---|
|  |  | ORN | Ctr[a] | ORN | Ctr[a] |
| **Age** |  |  |  |  |  |
| Median(range) | 62(11-93) | 58(46-77) | 61(14-93) | 60(46-83) | 65(11-91) |
| **Tumor Stage** [# of patients (% of patients)] |  |  |  |  |  |
| Stage I | 130(10.3) | 2(8.0) | 82(9.1) | 2(22.2) | 44(13.5) |
| Stage II | 151(11.9) | 2(8.0) | 102(11.3) | 0(0.0) | 47(14.4) |
| Stage III | 182(14.4) | 4(16.0) | 137(15.1) | 2(22.2) | 39(12.0) |
| Stage IV | 669(52.8) | 16(61.5) | 512(56.6) | 4(44.4) | 137(42.0) |
| Stage X (undefined) | 134(10.6) | 2(8.0) | 72(8.0) | 1(11.1) | 59(18.1) |
| **Gender** [# of patients (% of patients)] |  |  |  |  |  |
| Female | 327(25.8) | 6(23.1) | 244(27.0) | 1(11.1) | 76(23.3) |
| Male | 939(74.2) | 20(76.9) | 661(73.0) | 8(88.9) | 250(76.7) |
| **Concurrent Chemotherapy**[b] [# of patients (% of patients)] |  |  |  |  |  |
| w/ concurrent chemotherapy | 593(58.0) | 17(70.8) | 459(58.6) | 4(66.7) | 113(54.1) |
| **Smoker**[b] [# of patients (% of patients)] |  |  |  |  |  |
| Smoker | 533(52.1) | 12(50.0) | 443(56.5) | 5(83.3) | 73(34.9) |
| **Current Smoker[b]** [# of patients (% of patients)] | | | | | |
| Current smoker | 109(10.7) | 5(20.8) | 95(12.1) | 0(0.0) | 9(4.3) |
| **Hypertension[b]** [# of patients (% of patients)] | | | | | |
| w/ hypertension | 513(50.2) | 13(54.2) | 418(53.3) | 2(33.3) | 80(38.3) |
| **Diabetes[b]** [# of patients (% of patients)] | | | | | |
| w/ diabetes | 146(14.3) | 2(8.3) | 116(14.8) | 0(0.0) | 28(13.4) |
| **Dental Extraction[b]** [# of patients (% of patients)] | | | | | |
| w/ dental extraction | 167(16.3) | 4(16.7) | 121(15.4) | 2(33.3) | 40(19.1) |
| Patients | | | | | |
| **I 1[c]** (# of patients) | 1023 | 24 | 784 | 6 | 209 |
| **I 2[d]** (# of patients) | 243 | 2 | 121 | 3 | 117 |
| **Total** (# of patients) | 1266 | 26 | 905 | 9 | 326 |

[a]Ctr: Control group

[b]Data collected from Mayo Clinic in Rochester only

[c] I 1: Mayo Clinic in Rochester

[d] I 2: Mayo Clinic in Arizona

**Matching Method between VMAT and PBSPT group**

To make sure the comparison between two treatment modality groups and the RBE calculation was minimally impacted by the imbalance of clinical factors, we employed propensity score matching (PSM) to construct matched cohorts(11,44). Specifically, 335 patients were selected from the VMAT group to form a 1:1 matched study cohort with 335 patients from the PBSPT group(45). The matching process comprehensively considered multiple clinical factors, including age, tumor stage, gender, concurrent chemotherapy, current smoker, and smoking history, hypertension, diabetes, and dental extraction. The optimal matching strategy was adopted to identify paired patients that optimize overall matching quality, with each patient matched only once (44). For technical implementation, propensity score calculation and the matching process were completed using "MatchIt" package of R (version 4.4.1).

Regarding missing values present in the dataset, considering that the missing patterns exhibited structural characteristics, imputation methods were not employed. Instead, missing values were coded as independent categories and incorporated into the analysis as valid information for model training, thereby preserving data integrity and authenticity.

For the statistical analysis, categorical clinical factors—such as gender, tumor stage, smoking history, current smoker, chemotherapy, hypertension, diabetes, and dental extraction—were assessed using the chi-square test. Continuous clinical factors, such as age, were evaluated using a two-sided, two-sample t-test. Statistical significance was defined as a p-value less than 0.05.

**Within-Group Clinical Factor Balance Assessment**

It is important to note that, in addition to ensuring the balance of clinical factors between the VMAT and PBSPT groups, achieving balance **within** each group is also essential for obtaining unbiased results. Within-group balance of a clinical factor, conditional on a specific dosimetric factor, implies that the clinical factor is uniformly distributed across different levels of that dosimetric factor. To quantify this balance, we calculated standardized mean differences (SMDs). As an initial step, we considered the case with one clinical factor and one dosimetric factor. Specifically, the dataset was stratified into four subgroups according to the dosimetric factor values: high, mid-high, mid-low, and low. SMDs of the clinical factor were then computed across all

pairwise comparisons of these four subgroups. The SMD is defined as

$$\text{SMD}_{i,j} = \left|\frac{\mu_i - \mu_j}{\sigma_{pooled,ij}}\right|,$$

Where the $i$ and $j$ indicate different subgroups. $\mu_i$ and $\mu_j$ represent the mean values of the clinical factor in the two comparison subgroups $i$ and $j$. $\sigma_{pooled,ij}$ is the pooled standard deviation between subgroups, which is $\sigma_{pooled,ij} = \sqrt{(\sigma_i^2 + \sigma_j^2)/2}$. $\sigma_i$ and $\sigma_j$ are the standard deviation of the corresponding subgroups. We used the worst-case standardized mean differences (SMD) $\text{SMD}_{max} = \max_{i \neq j} \text{SMD}_{i,j}$ to quantify the within group balance of the clinical factor conditional on a specific dosimetric factor.

Thresholds were applied to interpret the extent of imbalance (46):

- $\text{SMD}_{max} < 0.2$: Considered well-balanced.

- $0.2 \leq \text{SMD}_{max} < 0.8$: Considered moderately biased.

- $\text{SMD}_{max} \geq 0.8$: Considered highly biased.

For our data with multiple clinical factors and multiple dosimetric factors, we separately calculated the SMD of each clinical factor conditional on each dosimetric factor to assess the within-group imbalance of all the clinical factors.

This analysis provided a quantitative overview of residual confounding bias inside the VMAT or PBSPT group after case-matching. For technical implementation, SMD calculations were completed using the in-house developed R (version 4.4.1) code.

**Causal effect and average treatment effect**

In causal effect analysis, patient-level variables are typically categorized into three groups. First are the **treatment variables**, represented in this study by dosimetric factors. Second are the **covariates**, which may influence treatment assignment, outcomes, or both; in our analysis, these correspond to clinical factors. Third is the **outcome variable**, which in this study is the incidence of osteoradionecrosis (ORN). The causal relation between dosimetric factors and ORN was quantified by the treatment effect. Here, we used the average treatment effect (ATE) to measure the causal relationship between the treatment and the potential outcomes in the population level(47):

$$ATE = \mathbb{E}[Y(T = 1) - Y(T = 0)]$$

where $Y(T = 1)$ and $Y(T = 0)$ are the potentially treated and controlled outcomes of the whole population. $T$ indicates the treatment variable. When the outcome was normalized, for binary treatments, ATE represented the probability of change in the outcome when applying the treatment compared to not applying it. In this work, T=0 or 1 represented whether the dosimetric factor value is below or above a certain threshold.

As shown in Figure 1(a), randomized controlled trials (RCTs) are the gold standard for analyzing the causal effects between treatment variables and patient outcomes. Through randomization, RCTs break the association between treatment assignment and patient clinical factors, making treatment assignment independent of clinical factors, thereby effectively controlling biases introduced by clinical factors in treatment variable assignment. However, in real-world data (RWD) that often contains significant biases, traditional statistical methods alone cannot capture the true causal effects(39). To calculate the causal relationship between the dosimetric factors and ORN from the biased real-world data, we used the causal inference approach in this work.

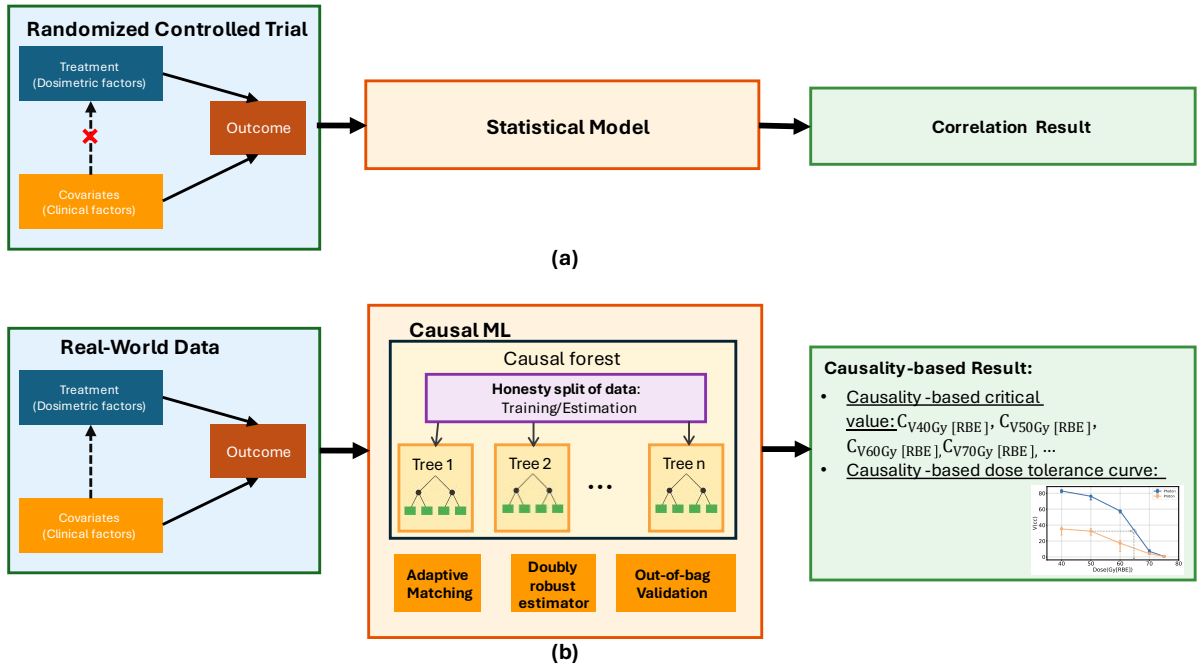

**Figure 1.** Two approaches to analyze causal effects. Randomized Controlled Trials (RCTs) eliminate bias from clinical factors by randomly assigning treatments, allowing standard statistical methods to infer causal effects between dosimetric factors and outcomes. However, the real-world data contains biased clinical factors. As a result, traditional statistical methods can only reveal correlations rather than true causal relationships from the real-word data. In this study, we employ causal forest (CF), a causal machine learning (Causal ML) approach. The CF enables us to estimate the causal effect of dosimetric factors upon patient outcome.

**Causal Forest**

In this study, we applied the **causal forest (CF)** method to estimate treatment effects, including the average treatment effect (ATE)(42,43,48). CF extends the traditional random forest framework to causal inference by leveraging the generalized random forest (GRF) approach, enabling flexible modeling of heterogeneous treatment effects. As an ensemble learning method, CF comprises a collection of decision trees. During training, each tree is fit on a random subsample of the dataset obtained through sampling without replacement. At each split, all potential cut points are evaluated to divide nodes into left and right child nodes. The key principle of CF is to maximize heterogeneity in treatment effects across child nodes by splitting on clinical factors that influence both treatment effects and treatment propensities. This strategy helps identify and adjust for biases introduced by clinical factors during causal effect estimation. To estimate the average treatment effect and corresponding standard errors, we employed the out-of-bag (OOB) approach, which leverages the entire sample without requiring an explicit train–test split (43). Additional technical details of the CF methodology are provided in the Supplementary Materials.

The CF method enables robust estimation of causal effects and ATE of dosimetric factors on ORN while effectively controls for biases inherent in clinical factors. The 'GRF-2.4.0' package of R (version 4.4.1) was utilized to construct the causal forest algorithm.

**Calculation of critical dose volume constraints**

DVCs were derived by identifying the thresholds that yield the maximum ATE. We binarized the dose-volume histogram (DVH) indices using a threshold value. The ATE of the binarized DVH indices was then calculated using CF. The derived critical volumes of the corresponding DVCs formed the volume tolerance curves(49).

For each dosimetric factor, we systematically evaluated different threshold values at 1% increments and computed their corresponding ATEs. The threshold associated with the most significant ATE was the critical dose-volume constraint. To ensure statistical robustness, 95% confidence intervals for these critical dose-volume constraints were obtained through 1,000 bootstrap iterations.

**Calculation of Empirical RBEs**

RBE is defined as the ratio of doses to achieve an identical clinical endpoint when comparing a new radiotherapy modality to conventional photon radiotherapy, like PBSPT to VMAT in this study. The empirical proton RBEs were derived by comparing the volume tolerance curves(49) between PBSPT and VMAT. In this study, the critical volumes of the derived DVCs were considered as the endpoints. We linearly interpolated the VMAT volume–tolerance curve to obtain equivalent constraint doses, such that, the VMAT critical volumes were equal to that of PBSPT for different physical doses. The empirical RBEs of PBSPT for mandibular osteoradionecrosis (ORN) were then computed as the ratio of the equivalent constraint dose to the corresponding PBSPT physical dose (RBE=1.0). Bootstrap analysis with 1,000 iterations provided 95% confidence intervals for the RBE calculations.

**Robustness check of Causal Machine Learning**

To ensure the reliability of the CF method, we conducted a robustness check. In this robustness check, we sequentially set the value of each normalized dosimetric factor to a random number between 0 and 1 uniformly and used these synthetic dosimetric data to repeatedly train the CF and calculate the corresponding ATE. If the ATE derived using these synthetic dosimetric data approaches zero, it indicated that the GRF method can effectively identify the ATE in real-world data. The 95% confidence intervals for the robustness check were calculated through 1,000 bootstrap iterations.

## Results

**Group Balance Before and After Case Matching**

Table 2 presents the p-values of clinical covariates comparing the VMAT and PBSPT groups before and after case matching. In the unmatched dataset (n = 1262), statistically significant differences ($p < 0.05$) were observed between the two groups across nearly all clinical factors, including tumor stage, concurrent chemotherapy, hypertension, diabetes, dental extraction, smoking history, and current smoking status – suggesting baseline imbalance between groups.

After applying the matching procedure (resulting in n = 670), the group differences were markedly reduced. None of the clinical factors showed statistically significant differences (all $p > 0.05$), suggesting that the matching process successfully improved clinical factor balancing between the two treatment groups. Notably, clinical factors such as tumor stage ($p < 0.001$ before vs. $p = 0.943$ after) and dental extraction ($p < 0.001$ before vs. $p = 0.669$ after) achieved substantial improvements in balance. Our dataset includes the subsites of oropharyngeal cancer and oral cavity cancer. After case-matching, the balance of subsites between the VMAT and PBSPT groups was also verified (p=0.064). These results demonstrate the effectiveness of the case-matching procedure in reducing confounding due to clinical covariates between VMAT and PBSPT groups.

**Table 2:** P-value between VMAT and PBSPT group before and after the case-matching.

| Metrics | Unmatched (VMAT=931; PBSPT=335) | Matched (VMAT=335; PBSPT=335) |
| --- | --- | --- |
| **Age** | 0.568 | 0.698 |
| **Tumor Stage** | <0.001 | 0.963 |
| **Gender** | 0.188 | 0.370 |
| **Concurrent Chemotherapy** | <0.001 | 0.498 |
| **Smoking History** | <0.001 | 0.765 |
| **Current Smoker** | <0.001 | 0.525 |
| **Hypertension** | <0.001 | 0.757 |
| **Diabetes** | <0.001 | 0.742 |
| **Dental Extraction** | <0.001 | 0.762 |

**Bias Checking of the clinical factors after case-matching**

The separately calculated SMD of each clinical factor conditional on each dosimetric factor for the (a) VMAT group and (b) PBSPT group after case matching are shown in Figure. 2. In the VMAT group, only the current smoker on V70Gy[RBE] are balanced. For the moderate dose volume indices, current smoker and diabetes on V40Gy[RBE] and V50Gy[RBE] were highly biased. For the high dose volume indices, gender and chemotherapy on V60Gy[RBE], and chemotherapy and smoking history on V70Gy[RBE] were highly biased. Chemotherapy and smoking history on $D_{mean}$ and all clinical factors except age and gender on $D_{max}$ were highly biased.

In the PBSPT group, all clinical factors were biased. Especially, clinical factors on dose volume indices V40Gy[RBE], V50Gy[RBE], V60Gy[RBE] and V70Gy[RBE], were more biased than the VMAT group. Chemotherapy and dental extraction on V60Gy[RBE] were highly biased. Only the age and tumor stage on V40Gy[RBE], V50Gy[RBE], and tumor stage and dental extraction on V70Gy[RBE] were moderately biased. The clinical factors of age, gender, chemotherapy, current smoker, diabetes, hypertension and dental extraction on $D_{max}$ were highly biased. Current smoker and diabetes on $D_{mean}$ were highly biased. Tumor stage was the only moderately biased clinical factor across all dosimetric factors in PBSPT group.

This analysis highlights that case-matching can only balance the clinical factors between VMAT and PBSPT groups. The residual biases inside each group were still significant. These residual biases underscore the necessity of adopting causal techniques—such as CF—to properly control the clinical factors and obtain the treatment effect of dosimetric factors on patient outcomes.

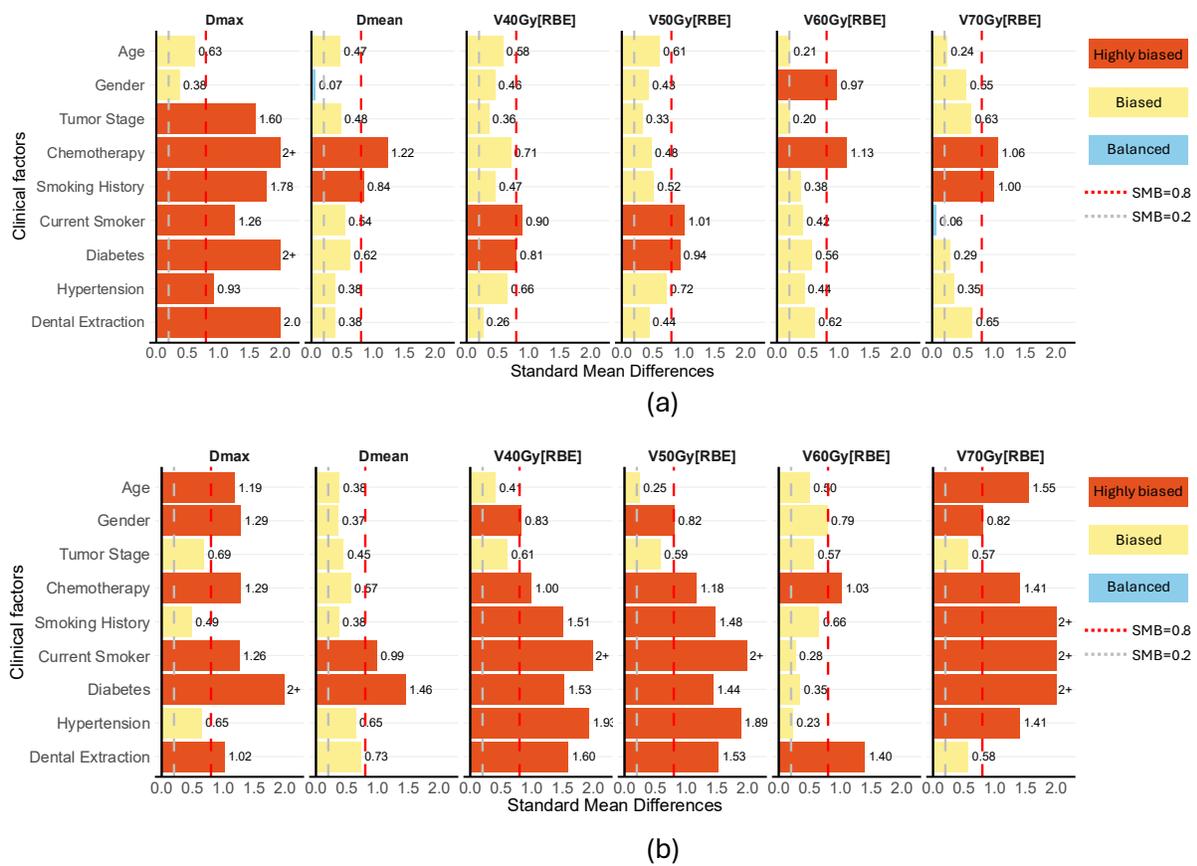

**Figure 2.** Standard Mean Differences (SMD) of clinical factors conditional on each dosimetric

factor for the (a) VMAT and (b) PBSPT groups after the case-matching between the VMAT and PBSPT groups. The vertical red dashed line represents the highly biased threshold at SMB=0.8. The boxes higher than this value are colored orange. The vertical gray dashed line represents the biased threshold at SMB=0.2. The boxes with 0.2<SMB<0.8 are colored yellow.

**Critical Volumes of dosimetric factors and robustness check**

The critical volumes of dosimetric factors were obtained and shown in Table 3. For the VMAT group, the critical volumes of the derived mandible DVCs are V40Gy[RBE]: 82.92cc (95% CI: 80.84cc -84.21 cc); V50Gy[RBE]: 76.25 cc (95% CI: 71.74–77.45 cc); V60Gy[RBE]: 57.77 cc (95% CI: 55.30–58.21 cc) and V70Gy[RBE]: 7.05 cc (95% CI: 5.77–8.96 cc); The critical value for $D_{max}$ was 7299.58 cGy[RBE] (95% CI: 1229.12–7596.32 cGy[RBE]), and the mean dose ($D_{mean}$) was 6131.54 cGy[RBE] (95% CI: 6080.88–6147.56 cGy[RBE]).

For the PBSPT group, the critical volumes were V40Gy[RBE]: 35.17 cc (95% CI: 27.62–37.45 cc); V50Gy[RBE]: 32.39 cc (95% CI: 27.36–35.55 cc); V60Gy[RBE]: 17.26 cc (95% CI: 6.85–20.74 cc) and V70Gy[RBE]: 4.00 cc (95% CI: 3.44–4.24 cc). The critical value for $D_{max}$ was 4339.49 cGy[RBE] (95% CI: 609.80–6524.71 cGy[RBE]), and the mean dose ($D_{mean}$) was 3282.88 cGy[RBE] (95% CI: 3247.35–3888.17 cGy[RBE]). VMAT has higher critical volumes than PBSPT for V40Gy[RBE], V50Gy[RBE], V60Gy[RBE] and V70Gy[RBE].

The robustness check and the 95% CI are shown in Supplemental Table 5. The maximum ATE in robustness check is lower than 0.02, which shows the consistency and robustness of the CF method on this dataset.

Table. 3 Critical volumes, average treatment effect (ATE), equivalent constraint dose in photon and empirical RBE for the VMAT and PBSPT

|  | V40Gy[RBE](cc) | V50Gy[RBE](cc) | V60Gy[RBE](cc) |
|---|---|---|---|
| **Critical volumes(cc) (95%CI)** | | | |
| Photon | 82.92(80.84-84.21) | 76.25(71.74- 77.45) | 57.77(55.30-58.21) |
| Proton | 35.17(27.62-37.45) | 32.38(27.36-35.55) | 17.26(6.85-20.74) |
| **ATE(95%CI)** | | | |
| Photon | 0.393(0.119-0.978) | 0.611(0.235-0.980) | 0.975(0.969-0.981) |
| Proton | 0.121(0.07-0.219) | 0.274(0.117-0.980) | 0.107(0.064-0.223) |
| **Equivalent constraint dose in photon (Gy[RBE=1.0]) (95%CI)** | 58.58(57.96-60.03) | 59.10(58.36-60.13) | 61.75(61.04-63.05) |
| **Empirical RBE (95% CI)** | 1.611(1.595-1.650) | 1.300(1.284-1.320) | 1.132(1.119-1.156) |

**Volume tolerance curve and Empirical RBEs**

Figure 3 illustrates the volume tolerance curves for both the VMAT and PBSPT groups. Empirical relative biological effectiveness (RBE) values were derived by comparing these curves through equivalent constraint dose analysis, as shown in Figure 3C. At proton doses of 40 Gy[RBE=1.1], 50 Gy[RBE=1.1], and 60 Gy[RBE=1.1], the corresponding empirical RBEs were 1.61 (95% CI: 1.60–1.65), 1.30 (95% CI: 1.28–1.32), and 1.13 (95% CI: 1.12–1.16), respectively (Table 3). Notably, the empirical RBE values exhibited a decreasing trend with increasing physical proton dose and close to 1.1 near 60 Gy[RBE=1.1] in the PBSPT group.

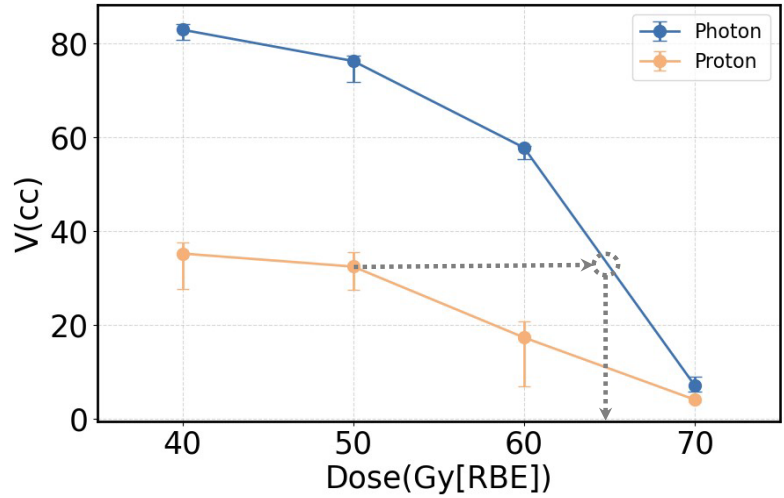

**Figure.3** Volume tolerance curves for VMAT (blue) and PBSPT (yellow) based on the derived DVCs. Gray circle indicates the position at the intersection between the VMAT volume tolerance curve and a horizontal line (the gray arrow) with the same critical volume value as the corresponding DVC of the PBSPT volume tolerance curve. Error bars indicate the 95% confidence intervals.

# Discussion

To the best of our knowledge, this is the first application of causal effect analysis to radiation oncology patient outcomes. We demonstrated the significant confounding bias from the clinical factors to the dosimetric factors. This significance implies that correlation results derived solely from traditional statistical approaches may not reliably reflect the causal relationship between dose and patient outcomes. By employing the Causal Forest (CF) method, we were able to identify causally informed critical values of dosimetric factors, leading to a more robust and unbiased RBE estimation.

The primary objective of this study was to investigate the causally informed empirical RBE associated with mandibular ORN in H&N cancer patients undergoing PBSPT. We included a total of 1,266 patients from our institution to report the incidence of ORN and to summarize all known clinical factors potentially associated with ORN in patients treated with either VMAT or PBSPT. This comprehensive dataset provides a broader and clinically relevant perspective on the occurrence of mandibular ORN in head and neck (H&N) cancer patients receiving these two radiotherapy modalities.

The VMAT and PBSPT groups were firstly balanced with respect to all known clinical risk factors for ORN. We applied propensity score matching to match 335 VMAT patients with 335 PBSPT patients (n = 670 in total), ensuring comparability across groups. After matching, the imbalance between the VMAT and PBSPT groups was substantially reduced, particularly for key clinical factors such as tumor stage, concurrent chemotherapy, hypertension, diabetes, dental extraction, and smoking history.

However, our bias analysis based on standardized mean differences (SMD) revealed that, even after achieving balance between the VMAT and PBSPT groups, significant imbalances remained within each group on clinical factors. If one were to use statistical methods—including traditional machine learning approaches such as support vector machines (SVMs)—to analyze the association between dosimetric variables and ORN under these conditions, the resulting associations would not reliably reflect causal relationships. As a result, conclusions drawn purely from such correlational analyses would be subject to unavoidable bias.

To obtain more reliable insights, it is crucial to consider **causal relationships** rather than mere correlations. Causality goes a step beyond correlation by identifying whether one factor truly influences another. For instance, dysphagia has been shown to correlate with radiation dose to the pharyngeal and laryngeal regions. However, several underlying scenarios could explain this association:(1) A high dose to the pharyngeal and laryngeal regions causes dysphagia, as is widely believed in the radiotherapy (RT) community; (2) Patients who present with dysphagia may, by coincidence (adjacent to primary tumor or involved lymph nodes) or selection bias (direct tumor involvement), receive higher doses to these regions (a case of reverse causation); (3) Both high dose and dysphagia may be influenced by a third factor—such as primary tumor location or smoking history—that confounds the observed relationship.

As illustrated in Figure 1, modern prospective clinical trials with careful design on randomization and control of potential confounding variables are the gold standard for causal inference analysis. However, they are often costly, time-consuming, and constrained in the number of treatment arms they can feasibly investigate. In contrast, modeling based on real-world observational data offers a more scalable and cost-effective approach. When combined with robust causal inference methods, such modeling could also provide causal insights, such as treatment effects.

To appropriately address bias in our dataset, we adopt the causal forest model to derive the critical values for dosimetric factors. The critical values correspond to points where the estimated ATE is most significant. In other words, when patients are stratified based on these critical values, the difference in toxicity risk between the low-dose and high-dose groups is both statistically significant and causally attributable to the dose itself, rather than to confounding variables. As a result, the derived DVCs not only capture key inflection points in the dose-toxicity relationship but also provide actionable guidance for future treatment planning.

Our results indicated that PBSPT patients had lower critical tolerance volumes compared to VMAT patients (Figure 3 and Table 3). This suggests that the RBEs for mandibular ORN in the 40–60 Gy [RBE = 1.1] dose range may be higher than 1.1. Notably, compared to our previous correlation-based study using the same patient cohort, the critical values identified through this causal analysis differed significantly, yet yielded similar RBE-related conclusions. This highlights that a substantial underestimation of RBE in the moderate dose range may contribute to unexpected mandibular ORN in head and neck cancer patients treated with PBSPT. However, the underlying mechanisms for RBEs exceeding 1.1 warrant further investigation.

Despite the valuable insights gained, several limitations warrant consideration. First, while we endeavored to satisfy the three key assumptions of causal inference (independence, ignitability, and positivity), achieving complete fulfillment would require even more comprehensive clinical databases. Second, it's important to recognize that RBE varies with clinical endpoints and tissue types; our study specifically investigated empirical RBE for mandibular ORN only. Given that ORN is a late-occurring complication, a study with longer follow-up duration might provide more definitive results. Finally, while our approach represents a methodological advancement, prospective validation in independent cohorts would further strengthen our findings.

## Conclusion

In this study, we introduce a novel application of causal machine learning to patient outcome analysis in radiotherapy. By identifying the critical values of DVH indices corresponding to the largest causal effects, we established a framework for deriving empirical RBEs through equivalent constraint dose analysis based on volume tolerance curves. Our findings suggest that the RBE for protons may substantially exceed 1.1 in the moderate dose range (40–60 Gy [RBE=1.1]), highlighting the need to consider this in future treatment planning to reduce the risk of osteoradionecrosis (ORN). This work demonstrates the feasibility and utility of causal machine learning in radiotherapy outcome studies, offering a robust complement to traditional statistical approaches.